\pdfoutput=1
\documentclass[11pt,twoside,a4paper,cmspaper,final,collab]{cms-tdr}
\def\svnVersion{d702f16}\def\svnDate{2023/07/24}\def\cmsCernNoTag{CERN-EP-2022-095}\def\cmsCernDate{\today}\def\cmsMessage{Published in the Journal of High Energy Physics as \href{http://dx.doi.org/10.1007/JHEP07(2023)148}{\doi{10.1007/JHEP07(2023)148}.}}
\begin{document}\cmsNoteHeader{HIG-21-003}

\newcommand{\sqrts}{\ensuremath{\sqrt{s}=13TeV}\xspace}
\newcommand{\hgggg}{\ensuremath{\PH \to  aa \to \PGg\PGg\PGg\PGg}\xspace}
\newcommand{\haa}{\ensuremath{\PH \to  aa}\xspace}
\newcommand{\sigmaH}{\ensuremath{\sigma_{\PH}\xspace }}
\newcommand{\BR}{\ensuremath{\mathcal{B}(\hgggg)}\xspace}
\newcommand{\mgggg}{\ensuremath{m_{\PGg\PGg\PGg\PGg}}\xspace}
\newcommand{\dNLL}{\ensuremath{-2\Delta\ln{\text{L}}}\xspace}
\newcommand{\ptaone}{\ensuremath{p_{\mathrm{T},a1}}\xspace}
\newcommand{\ptatwo}{\ensuremath{p_{\mathrm{T},a2}}\xspace}
\newcommand{\DRa}{\ensuremath{\Delta R_{a_{1}a_{2}}}\xspace}
\newcommand{\Zee}{\ensuremath{\PZ \to \Pe\Pe }\xspace}
\newcommand{\GAMJETS}{\ensuremath{\PGg+\text{jets}}\xspace}
\newcommand{\DIPHO}{\ensuremath{\PGg\PGg+\text{jets}}\xspace}
\newcommand{\msc}{\ensuremath{m_{a}}\xspace}
\newcommand{\maone}{\ensuremath{m_{a1}}\xspace}
\newcommand{\matwo}{\ensuremath{m_{a2}}\xspace}
\newcommand{\mhyp}{\ensuremath{m_{a,\text{hyp}}\xspace}}

\newlength\cmsTabSkip\setlength{\cmsTabSkip}{1ex}

\providecommand{\cmsTable}[1]{\resizebox{\textwidth}{!}{#1}}
\providecommand{\cmsTableAlt}[1]{\resizebox{!}{0.7\textwidth}{#1}}

\cmsNoteHeader{HIG-21-003}

\title{Search for the exotic decay of the Higgs boson into two light pseudoscalars with four photons in the final state in proton-proton collisions at \texorpdfstring{$\sqrt{s} = 13\TeV$}{sqrt(s) = 13 TeV}}

\date{\today}

\author*[cern]{The CMS Collaboration}

\abstract{
A search for the exotic decay of the Higgs boson to a pair of light pseudoscalars, each of which subsequently decays into a pair of photons, is presented. The search uses data from proton-proton collisions at \texorpdfstring{$\sqrt{s} = 13\TeV$}{sqrt(s) = 13 TeV} recorded with the CMS detector at the LHC that corresponds to an integrated luminosity of 132\fbinv. The analysis probes pseudoscalar bosons with masses in the range 15--62\GeV, coming from the Higgs boson decay, which leads to four well-isolated photons in the final state. No significant deviation from the background-only hypothesis is observed. Upper limits are set on the product of the Higgs boson production cross section and branching fraction into four photons. The observed (expected) limits range from 0.80 (1.00)\unit{fb} for a pseudoscalar boson mass of 15\GeV to 0.26 (0.24)\unit{fb} for a mass of 62\GeV at 95\% confidence level.
}

\hypersetup{
pdfauthor={CMS Collaboration},
pdftitle={Search for exotic decay of the Higgs Boson into two light pseudoscalars with photons in the final state},%
pdfsubject={CMS},
pdfkeywords={CMS, Higgs}}

\maketitle

\section{Introduction}
\label{sec:intro}

In 2012, a Higgs boson (\PH) with a mass of 125\GeV was discovered by the ATLAS and CMS experiments~\cite{HiggsdiscoveryAtlas, Chatrchyan:2012ufa, Chatrchyan:2013lba} at the CERN LHC. Since its discovery, both collaborations have performed precision measurements of the spin, parity, width, and couplings of the Higgs boson in its various production and decay modes~\cite{Hgg_properties_13_v2,Hzz_xsec_13,Hzz_xsec_13_v2,ttH_13,ttH_13_v2,Hww_diff_13,CMS_mass,ATLAS_mass,ATLAS_xsec,Aaboud:2017vzb,Khachatryan:2016vau}, all of which indicate that Higgs boson properties are compatible with the standard model (SM) predictions. However, data collected at ${\sqrt{s}=13\TeV}$ provide an upper limit on the branching fraction of the Higgs boson to undetected states of about 40\% at 95\% confidence level (\CL)~\cite{Sirunyan:2018koj}. This leaves a large margin for beyond-the-SM (BSM) decays of the Higgs boson. Various theoretical models, such as 2HDM+S models~\cite{Curtin:2013fra}, describe a BSM Higgs boson that decays to light bosons, which has not yet been excluded.

Multiple searches for exotic decays of the Higgs boson have been performed using the 8\TeV~\cite{Aad:2015oqa,Khachatryan:2015wka,Khachatryan:2015nba,Khachatryan:2017mnf} and 13\TeV~\cite{Aaboud:2016oyb,Aaboud:2018fvk,Aaboud:2018gmx,Aaboud:2018iil,Aaboud:2018esj,Sirunyan:2018pzn,Sirunyan:2018mbx} data collected at the LHC. Decays of the type \haa, where \textit{a} is a light pseudoscalar boson, are well motivated in various BSM scenarios ~\cite{Chang:2006bw,Dobrescu:2000jt,Larios:2002ha,Dermisek:2007yt}, in particular in 2HDM+S models. In many scenarios, such as fermiophobic \textit{a} decays, the branching fraction of the pseudoscalar bosons to a pair of photons is close to unity, which enables this search at the LHC. The final state, with four photons, provides an experimental signature that has very small contributions from SM processes and is thus an important channel for the search for light pseudoscalar bosons.

\begin{figure*}[thb]
  \centering
  \includegraphics[width=.45\linewidth]{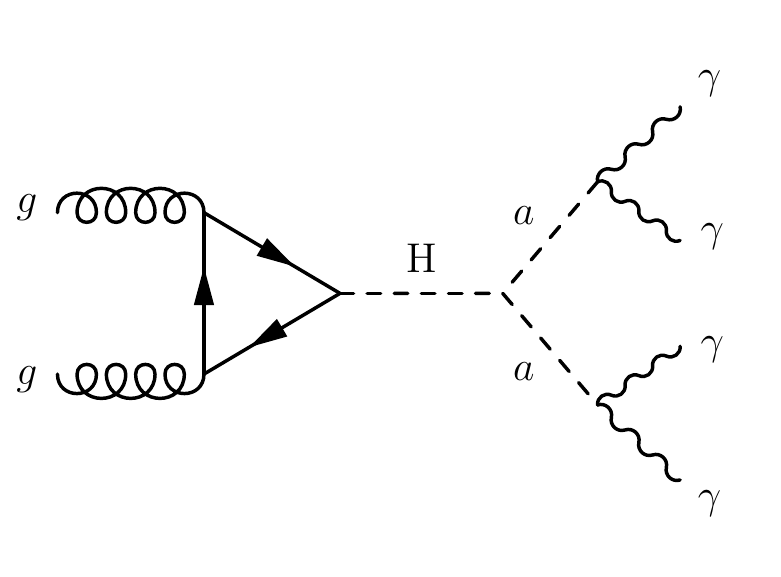}
  \caption{Feynman diagram for a BSM decay of the Higgs boson into a pair of light pseudoscalar bosons that subsequently decay into photons.}%
  \label{fig:feynman_h4g}
\end{figure*}

This paper presents a search for light pseudoscalar bosons that arise from the decay of a Higgs boson, with four photons in the final state: \hgggg. The event topology depends on the mass of the pseudoscalar boson being probed, which determines the opening angle between the photons for each pair. This analysis considers only events with four isolated photons in which the angular distance between the photons, for both photon pairs, is greater than 0.2. These requirements enable a search for pseudoscalar bosons that range in mass from 15 to 62\GeV and produce an experimental signature with four well isolated and fully reconstructed photons. The 15\GeV mass boundary is chosen because below that value most of the events have at least one merged photon pair or one photon that does not pass the reconstruction criteria. The signal reconstruction efficiency increases as the mass increases. The dominant Feynman diagram contributing to this process is shown in Fig.~\ref{fig:feynman_h4g}.

A previous search for light pseudoscalar bosons in events with at least three photons was performed by the ATLAS Collaboration using the LHC data collected at ${\sqrt{s}=8\TeV}$~\cite{ATLAS:2015rsn}. The first search of this type done by the CMS Collaboration is presented in this paper. Data collected by the CMS experiment from 2016 to 2018, which correspond to an integrated luminosity of 132\fbinv, are used in this analysis.

This paper is organized as follows. A brief description of the CMS detector is given in Section~\ref{sec:Detector}. Details of the data and simulation used in this analysis are described in Section~\ref{sec:Samples}. The reconstruction of \hgggg events and the associated selection requirements are summarized in Section~\ref{sec:EventReconstruction}. An overview of the analysis strategy is given in Section~\ref{sec:Strategy}. Section~\ref{sec:BkgEstimation} describes the background estimation technique, which is used as input to a multivariate discriminator. The optimization procedure, which uses the output of this discriminator, is discussed in Section~\ref{sec:Classification}. Sections~\ref{sec:StatProd},~\ref{sec:Signal} and ~\ref{sec:Background} describe the statistical procedure and the signal and background modeling, respectively. The systematic uncertainties are discussed in Section~\ref{sec:Systematics}. Finally, the results are presented in Section~\ref{sec:Results}, and the paper is summarized in Section~\ref{sec:Summary}.

\section{The CMS detector}
\label{sec:Detector}

The central feature of the CMS apparatus is a superconducting solenoid of 6\unit{m} internal diameter, providing a magnetic field of 3.8\unit{T}. Within the solenoid volume is a silicon pixel and strip tracker, a lead tungstate crystal electromagnetic calorimeter (ECAL), and a brass and scintillator hadron calorimeter (HCAL), each composed of a barrel and two endcap sections. Forward calorimeters extend the pseudorapidity ($\eta$) coverage provided by the barrel and endcap detectors. Muons are detected in gas-ionization chambers embedded in the steel flux-return yoke outside the solenoid.

Events of interest are selected using a two-tiered trigger system~\cite{Khachatryan:2016bia}. The first level, composed of custom hardware processors,
uses information from the calorimeters and muon detectors to select events at a rate of around 100\unit{kHz} within a time interval of less than 4\mus~\cite{Sirunyan:2020zal}. The second level, known as the high-level trigger, consists of a farm of processors running a version of the full event reconstruction software optimized for fast processing, and reduces the event rate to around 1\unit{kHz} before data storage~\cite{CMS_HLT}.

The particle-flow algorithm~\cite{CMS-PRF-14-001} aims to reconstruct and identify each individual particle in an event, with an optimized combination of information from the various elements of the CMS detector. The energy of photons is measured using the ECAL, as described in Ref.~\cite{CMS:2020ume}. In the barrel section of the ECAL, an energy resolution of about 1\% is achieved for unconverted or late-converting photons in the tens of GeV energy range. The energy resolution of the remaining barrel photons is about 1.3\% up to ${\abs{\eta} = 1}$, changing to about 2.5\% at ${\abs{\eta} = 1.4}$. In the endcaps, the energy resolution is about 2.5\% for unconverted or late-converting photons, and between 3 and 4\% for the photons that convert closer to the beam spot~\cite{CMS:2020ume,CMS:EGM-14-001}.

A more detailed description of the CMS detector, together with a definition of the coordinate system used and the relevant kinematic variables, can be found in Ref.~\cite{CMS:2008xjf}.

\section{Data samples and simulated events}
\label{sec:Samples}

The proton-proton (\Pp-\Pp) collision data at ${\sqrt{s}=13\TeV}$ used in this analysis were collected in 2016, 2017, and 2018 and correspond, respectively, to integrated luminosities of 36.3, 41.5, and 54.4\fbinv. This is 5.4\fbinv less than the standard CMS detector collected luminosity~\cite{CMS-PAS-LUM-17-001,CMS-PAS-LUM-17-004,CMS-PAS-LUM-18-002}, because a required trigger was missing for a short period.
Events are selected using a high-level diphoton trigger, optimized for the low-mass diphoton Higgs boson search~\cite{CMS:2018cyk}, with photon transverse momentum (\pt) thresholds of 30\GeV and 18\GeV. Additionally, calorimeter-based identification requirements, which use information such as the shape of the electromagnetic shower, the isolation of the photon candidate, and the ratio between the hadronic and electromagnetic energy deposit of the shower, are applied to the photon candidates at trigger level. Furthermore, the chosen trigger requires the diphoton candidate to have an invariant mass greater than 55\GeV in data collected during {2016--2017}. Each event is required to contain at least one diphoton candidate that satisfies these high-level trigger requirements. The invariant mass selection on the two highest \pt photons discards less than 25\% of the signal events. 

The simulated signal samples were generated corresponding to a pseudoscalar boson mass, \msc, ranging from 15 to 60\GeV, in steps of 5\GeV, assuming a Higgs boson mass of 125\GeV. These samples were simulated considering only the gluon fusion production mode of the Higgs boson, using \MGvATNLO~v2.4.2~\cite{madgraph}.

The dominant backgrounds in this search are SM production of \DIPHO, \GAMJETS, as well as multijet events, in which jets are misidentified as isolated photons. As in Refs.~\cite{Khachatryan:2014ira,Hgg_properties_13}, the background contributions are modeled entirely from data.

All simulated samples used in this analysis model QCD showering and hadronization with the \PYTHIA~8.212~\cite{Sjostrand:2014zea} event generator. The CUETP8M1 tune was used for data collected in 2016 and the CP5 tune was used for data collected in {2017--2018}~\cite{Khachatryan:2015pea, Sirunyan:2019dfx}. The response of the CMS detector is modeled using the \GEANTfour~\cite{GEANT4:2002zbu} package. The simulated events also include additional \Pp-\Pp interactions within the same or nearby bunch crossings (pileup). The average pileup in the 2016 ({2017--2018}) datasets is 23 (32) vertices. The simulated events are weighted to reproduce the distribution of pileup in data.

\section{Event reconstruction}
\label{sec:EventReconstruction}

The particle-flow algorithm~\cite{CMS-PRF-14-001} is used to reconstruct photon candidates from energy clusters in the ECAL that are not matched to any charged particle trajectories originating in the pixel detector. The energy of the photon candidates is calculated by applying in-situ measured calibrations to the reconstructed hits in the ECAL. A multivariate regression technique is employed to correct the photon energies measured by the ECAL. These procedures are described in Ref.~\cite{CMS:2020ume}.

Deposits from quark fragmentation and hadronization are clustered into hadronic jets. The energy of charged hadrons is determined from a combination of their momentum measured in the tracker and the matching ECAL and HCAL energy deposits, while the energy of neutral hadrons is obtained from the corresponding corrected ECAL and HCAL energy deposits. Jets are clustered using the anti-\kt jet finding algorithm~\cite{AntiKt,Cacciari:2011ma} with a distance parameter of 0.4. The missing transverse momentum vector \ptvecmiss is computed as the negative vector sum of the transverse momenta of all the PF candidates in an event, and its magnitude is denoted as \ptmiss~\cite{CMS:2019ctu}.

Prompt photons are distinguished from jets, which could be misidentified as a photon, using  a multivariate analysis (MVA) technique that uses information related to the photon's electromagnetic shower shape, isolation, energy, and $\eta$. This technique is detailed in Refs.~\cite{Hgg_properties_13,CMS_mass}.

Since photons do not leave deposits in the tracker, the most probable primary interaction vertex in the event is identified using a boosted decision tree (BDT). The primary vertex BDT is trained on simulated \hgggg events and uses input variables related to tracks recoiling against the four-photon system and information related to photons converted in the tracker material, similar to Ref.~\cite{Hgg_properties_13_v2}. A separate training is performed for each data-taking year ({2016--2018}) to properly model the variation in detector conditions over the three years, in particular with respect to the pixel detector upgrade. The analysis identifies the vertex with the highest BDT score as the primary vertex, which improves the resolution of the invariant mass of the Higgs boson candidate by approximately 3\%. It also increases the vertex identification efficiency, 80\% in total, which is defined as selecting a vertex within 1\unit{\cm} of the true generator vertex, by about 10\%, with respect to the vertex chosen with the largest value of summed $\pt^{2}$ of the tracks.

\section{Analysis strategy and selection}
\label{sec:Strategy}

Events considered in this analysis are required to contain at least one diphoton candidate where both daughter photon candidates pass the identification criteria, which are slightly more stringent than the trigger requirements. Additionally, events must contain at least four identified photon candidates in the ECAL and tracker fiducial region (${\abs{\eta} < 2.5}$). This excludes the ECAL barrel-endcap transition region (${1.44 < \abs{\eta} < 1.57}$) where the photon reconstruction is suboptimal. The four photon candidates are also required to pass \pt requirements. The \pt thresholds on the first- and second-leading photons are 30 and 18\GeV, respectively. These selections are in sync with with the \pt requirements at trigger level for the two leading photons. The thresholds on the third- and fourth-leading photons are 15\GeV, since the BDT-based photon identification criteria are optimal for ${\pt > 15}$\GeV. When more than four photon candidates are found, the four candidates with the highest \pt are chosen. Each photon candidate must pass an electron veto, based on the presence of geometrically compatible hits in the pixel detector. The four photon candidates with the highest \pt are also required to have an invariant mass (\mgggg) between 110 and 180\GeV. The lower bound is chosen to avoid \mgggg spectrum biases coming from the trigger selections. The Higgs boson candidate is constructed from the four photon candidates, which have passed all the previously described selection requirements.

To improve the sensitivity of the search, a 4-photon event classifier is trained to separate signal events from background events. The 4-photon event classifier utilizes the identification and kinematic information of the photons and pseudoscalar boson candidates. An optimized selection on the output of the event classifier is used to define the final signal regions for the analysis. The input variables for the classifier are uncorrelated with \mgggg; therefore, the shape of the \mgggg spectrum is not affected by any selections placed on the 4-photon event classifier, as verified in simulation.

When all four photons from the decay of the pseudoscalar boson pair are within the acceptance criteria of the analysis, the \hgggg signal will create a resonance in the \mgggg distribution at 125\GeV. The analysis performs an unbinned maximum likelihood fit of the signal and background models to the observed \mgggg distribution in data after a selection on the classifier output is applied. The signal model is constructed from a parametric fit to the simulated signal, while the background model is created using a data driven approach. After applying the 4-photon event classifier selection, the full analysis acceptance has a negligible dependence on the generated Higgs boson \pt or the non-gluon-fusion Higgs boson production modes. Therefore limits are set on the product of the Higgs boson inclusive production cross section and branching fraction into the four photon final state.

In this paper, ``nominal signal hypothesis" refers to simulated signal samples corresponding to a particular \msc value. This assumes a branching ratio of ${a \to \gamma\gamma}$ equal to unity. In these hypotheses, \msc ranges from {15--60}\GeV in steps of 5\GeV. The final results are reported with an \msc granularity of 0.5\GeV up to ${\msc = 40}$\GeV and 1\GeV for ${\msc > 40}$\GeV, where the signal models for the intermediate mass hypotheses are constructed using the resolution of the signal model of the closest nominal mass and by interpolating the signal model normalization between the nominal masses.

\section{Classifier training background}
\label{sec:BkgEstimation}

Because of the low event selection efficiency on the background samples, it is difficult to model the background from simulation. Therefore, the expected background model, which is used to train the 4-photon event classifier, is estimated from data.

The method, referred to as event mixing (a simplified version of ``Hemisphere mixing"~\cite{EventMixing}), does not rely on a control or sideband region, but instead aims to estimate the background contribution using the original dataset as input. This procedure begins with using data events that have passed the trigger selections, as described in Section~\ref{sec:Samples}, and replacing three out of the four reconstructed photons in each event with reconstructed photons from three consecutive events to create a ``mixed'' dataset. Photon candidates in the mixed dataset are required to pass the selections described in Section~\ref{sec:Strategy}. The psedoscalar boson candidates, a$_1$ and a$_2$, are reconstructed considering all possible pair combinations among the chosen four photons, and taking the two pairs with the smallest value of the difference between the pair invariant masses. The shuffling of the reconstructed photons between the events not only constructs a dataset that is similar to the original data, but is also insensitive to the possible presence of a resonant signal. This procedure is performed separately for data collected during 2016, 2017, and 2018.

{\tolerance=1200
The events in the mixed dataset can potentially have different kinematic properties from those in the original data. To correct this, a multi-dimensional per-event weight is calculated by comparing events from the mixed dataset to the original data in the \mgggg sideband region, i.e. {110--115} or {135--180}\GeV. The weight is computed from the ratio of the four-dimensional histograms filled with data and event mixing events, using the following variables: angular distance between the two pseudoscalar boson candidates, defined by $\Delta R_{a_{1}a_{2}} = \sqrt{\smash[b]{(\eta_{a1} - \eta_{a2})^{2} + (\phi_{a1} - \phi_{a2})^{2}}}$ where $\phi$ is azimuthal angle; the transverse momenta of the two pseudoscalar bosons (\ptaone and \ptatwo); and the difference between invariant mass of the two pseudoscalar boson candidates (${\maone-\matwo}$). This weight is applied to each event in the mixed dataset, and the reweighted events are then used to train the classifier. 
\par}

The event mixing dataset is used only for training the 4-photon classifier and to optimize the selection on the classifier score. Since the background model used in the final maximum likelihood fit is obtained directly from data, any residual disagreement between data and event mixing in the background-like regions cannot induce any bias, and it could only result in suboptimal performance of the classifier.

\section{Event classification}
\label{sec:Classification}

A dedicated 4-photon boosted decision tree (BDT) event classifier is trained to separate signal from background. The training sample is parameterized as a function of \msc in order to make the classifier output uniform and sensitive to the full range of signal hypotheses considered in the search~\cite{Baldi:2016fzo}. In this approach, a parameter equal to the hypothesized pseudoscalar boson mass (\mhyp) is provided as input to the training. The set of variables is chosen such that \mgggg cannot be inferred from the inputs. This is done by verifying that their correlation with \mgggg variable is negligible and that the \mgggg spectrum is not distorted by applying a selection on the classifier output.

The parameterized classifier requires only a single training and is able to provide a smooth interpolation to pseudoscalar boson mass hypotheses not used for the training. The training signal sample is obtained by merging all generated signal samples of equal size with masses between 15 and 60\GeV in steps of 5\GeV. The value of the \mhyp parameter is equal to the corresponding true mass for the nominal signal simulation. The event mixing dataset, as described in Section~\ref{sec:BkgEstimation}, is used as the background in the training. For the background, the value of the parameter \mhyp is randomly distributed as a flat function among all possible nominal \msc values in 5\GeV steps.

The variables used in the training help separate isolated photons originating from the decay of the pseudoscalar boson from those from prompt and non-prompt processes.
Pseudoscalar boson candidates are constructed from the four photon candidates as follows. For every possible combination of two photon candidate pairs, the difference between the invariant masses of the paired photons (${\Delta m}$) is evaluated. The pairs with the smallest value of ${\Delta m}$ are chosen to reconstruct the pseudoscalar boson candidates.

The following discriminating variables are provided as input to the training:

\begin{enumerate}
    \item Photon identification BDT score for all four photons.
    \item \pt of the two pseudoscalar boson candidates, \ie, \ptaone and  \ptatwo.
    \item Difference between the reconstructed invariant mass of the pseudoscalar boson candidates, \ie, ${\maone-\matwo}$.
    \item Difference between the invariant masses of the pseudoscalar boson candidate and the \mhyp parameter divided by \mgggg, \ie, ${(\maone-\mhyp)/\mgggg}$ and ${(\matwo-\mhyp)/\mgggg}$).
    \item Angular distance \DRa divided by \mgggg, \ie, ${\DRa/\mgggg}$.
    \item Angle ${\cos\theta_{a\PGg}^{*}}$ in the pseudoscalar boson rest frame, between the leading pseudoscalar boson candidate and the leading photon produced from its decay, chosen in the laboratory frame. This variable is sensitive to the spin of the pseudoscalar boson object.
\end{enumerate}

As part of this training procedure, simulated signal and background datasets from the three data-taking years ({2016--2018}) are scaled by their appropriate integrated luminosities and combined. This combination of datasets from three years provides large training statistics.
Additionally, the signal and background samples are divided in half to create the training and testing samples.

\begin{figure}[thbp]
  \centering
  \includegraphics[width=0.49\textwidth]{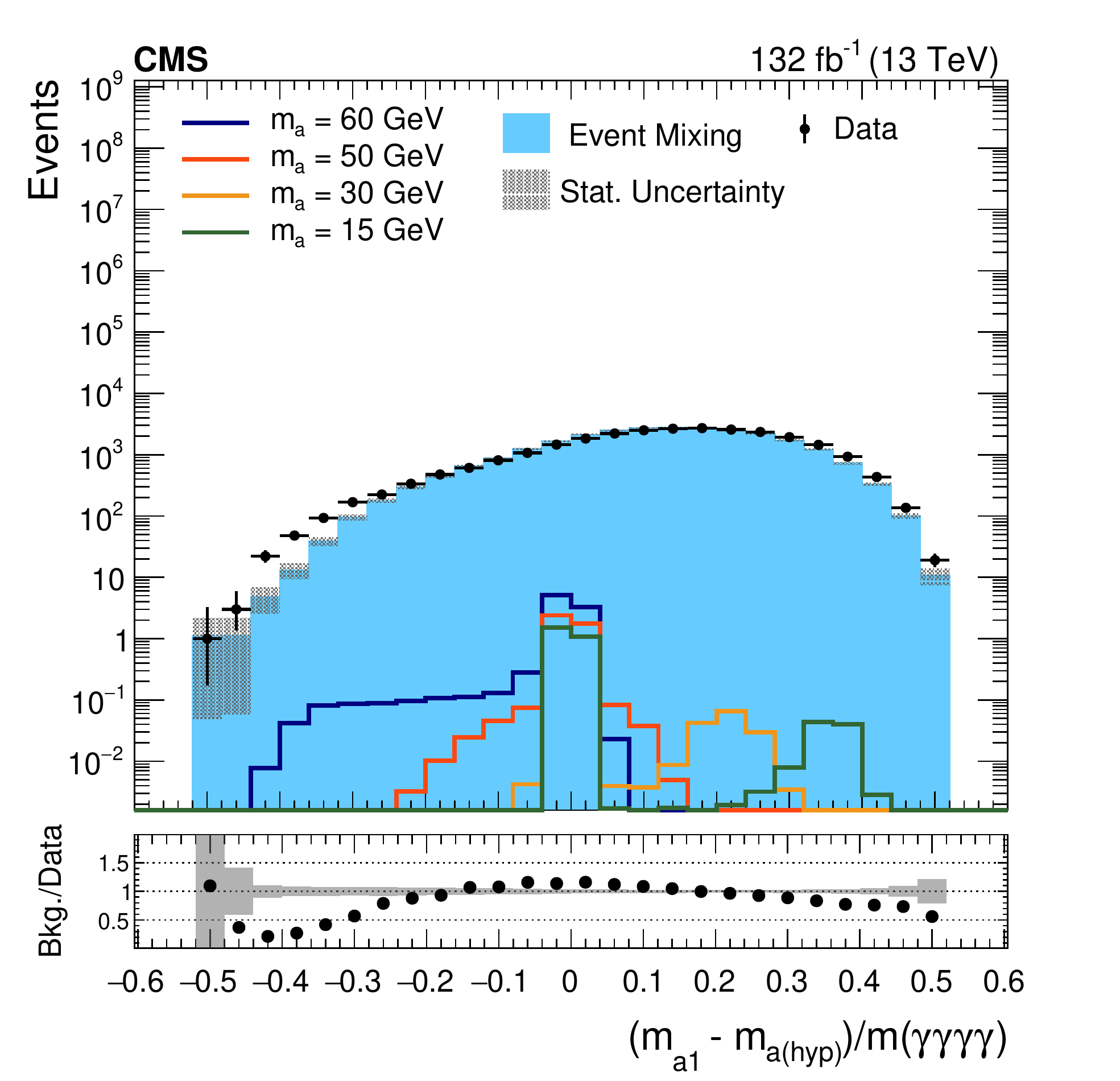}
  \includegraphics[width=0.49\textwidth]{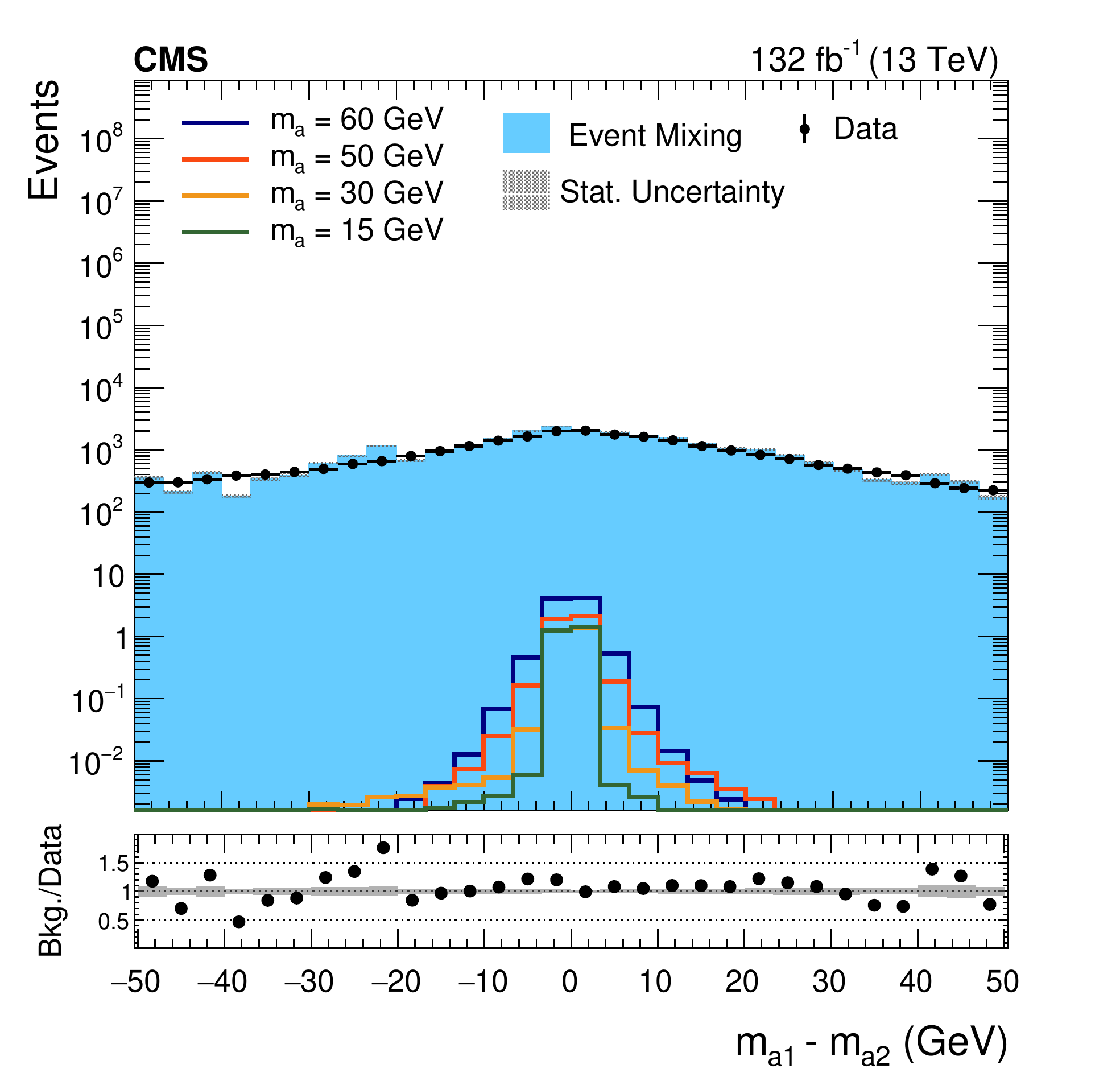}
  \includegraphics[width=0.49\textwidth]{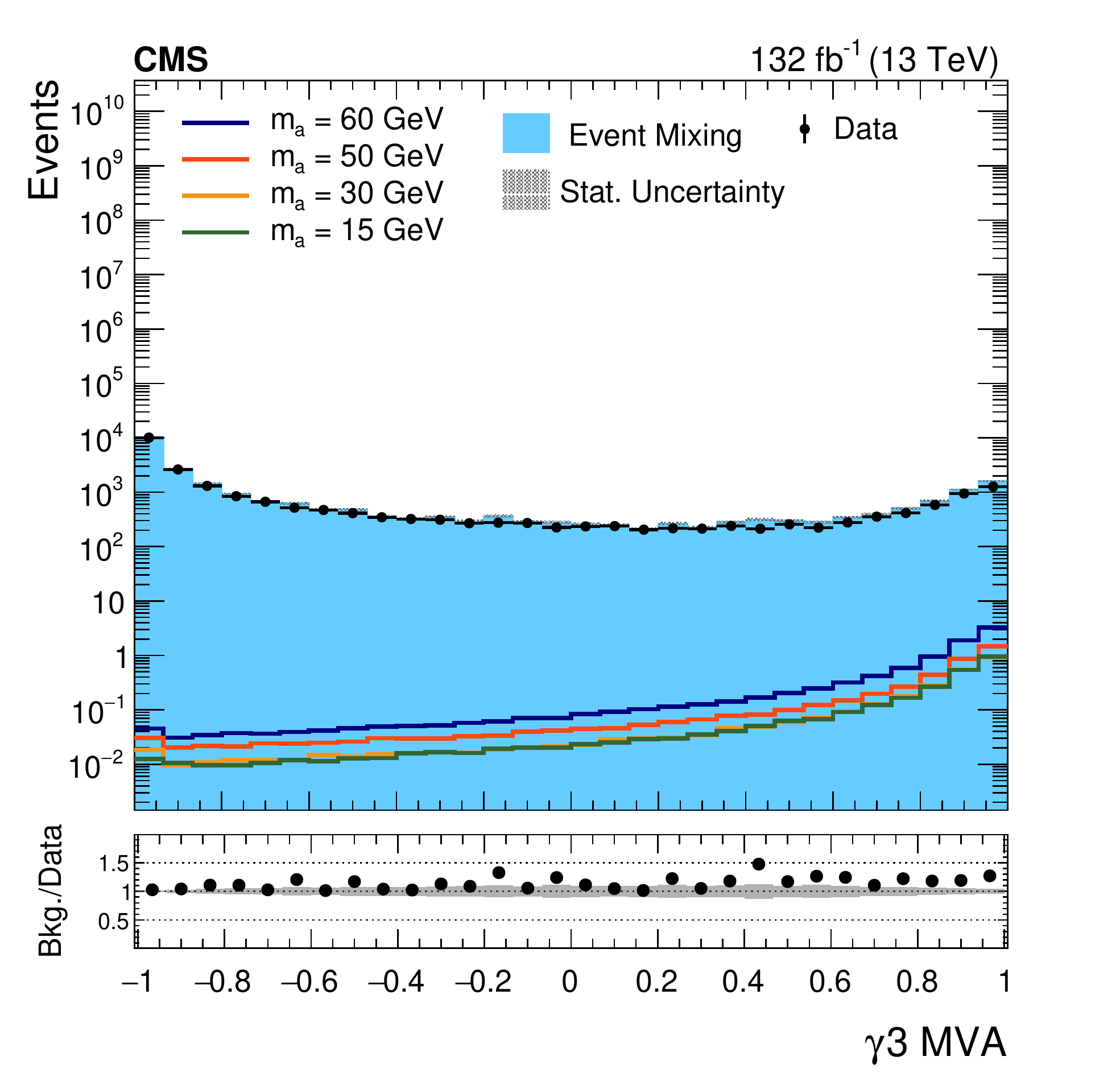}
  \includegraphics[width=0.49\textwidth]{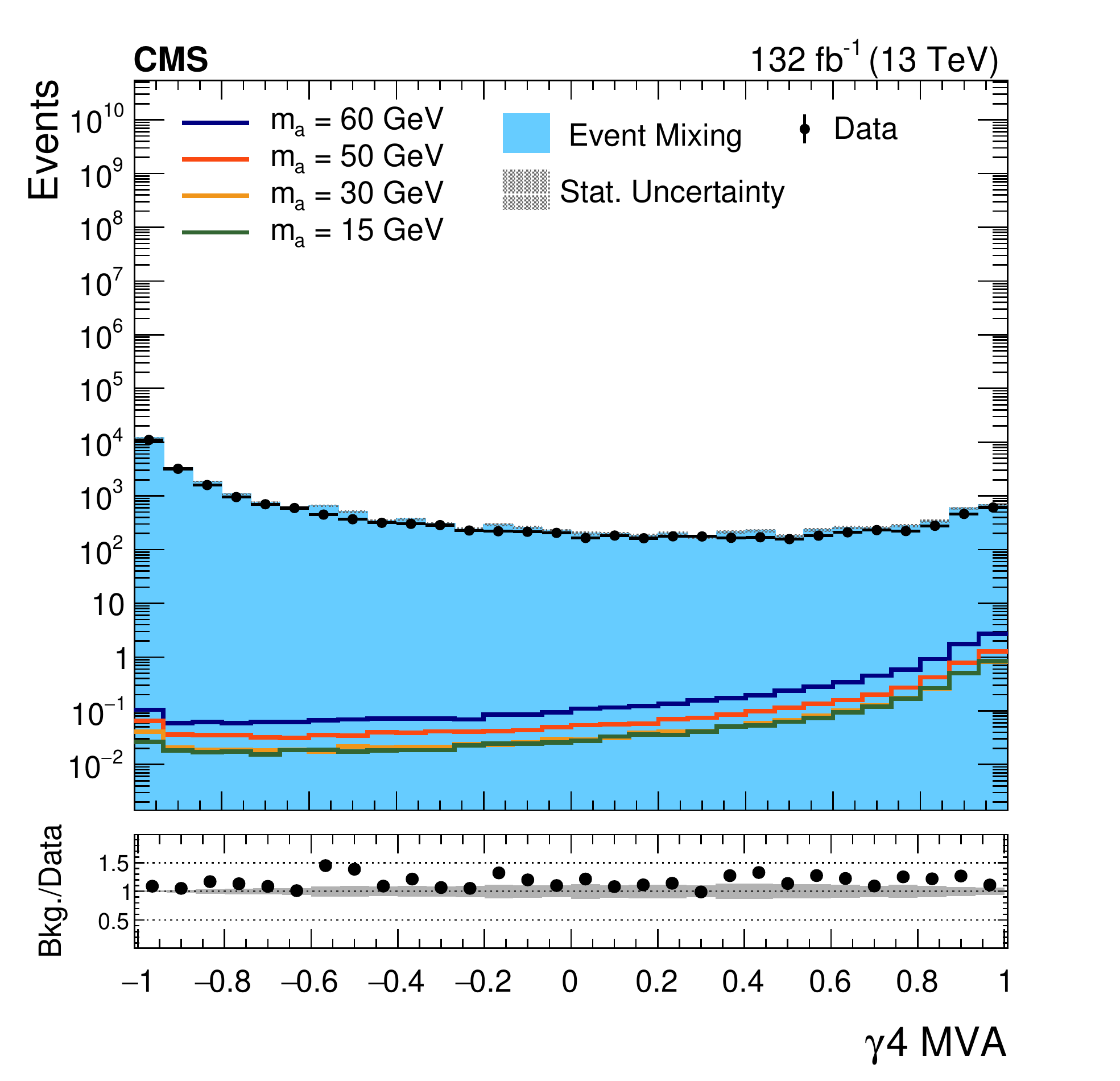}
  \caption{Distributions of the four most highly ranked discriminating variables: the difference between the invariant masses of the pseudoscalar bosons and the \mhyp parameter, divided by the invariant mass of the four-photon system (upper left), with off-zero signal peaks from photon pairing mismatches (those events are discarded after the 4-photon event classifier selection); the difference between the invariant masses of the pseudoscalar boson (upper right); the photon identification BDT score of the third leading, $\gamma_{3}$ (lower left) and the fourth leading, $\gamma_{4}$ (lower right) photons. The events shown are selected from the \mgggg sidebands (${110 < \mgggg < 115}$\GeV or ${135 < \mgggg < 180}$\GeV) for event mixing and data after fulfilling the selection criteria described in Section~\ref{sec:EventReconstruction}, while the signals are scaled with a cross-section of 1\unit{pb}.}
  \label{fig:InputVars}
\end{figure}

The distributions of the four most highly ranked variables in the training: ${(\maone-\mhyp)/\mgggg}$, ${\maone-\matwo}$, and the photon identification BDT score of the third and fourth photon are shown in Fig.~\ref{fig:InputVars} for the event mixing dataset, data, and signal simulation for various pseudoscalar boson mass hypotheses. It shows the contributions from the event mixing dataset and the data from the \mgggg sidebands, satisfying either $\mgggg = 110$--115 or 135--180\GeV. The distributions of the event mixing dataset and data are found to be in reasonable agreement, and the residual disagreement does not induce any biases in the analysis since the final background model is derived directly from data.

A unique 4-photon BDT output is obtained for each pseudoscalar boson mass hypothesis. The difference between the correlations of the input variables used in the training leads to a disagreement in the output shape of the BDT between the event mixing model and data. This difference is addressed by reweighting the BDT output distribution for the event mixing model to match output distribution for data from the \mgggg sideband region.

In order to maximize the sensitivity of the analysis, events are categorized according to the output of the 4-photon BDT. The categorization is optimized by maximizing the approximate mean significance (AMS)~\cite{Cowan:2010js} over all possible categories in a window covering the region: ${115 < \mgggg < 135}$\GeV. In particular, AMS is computed for each category, and the total AMS, defined as the sum in quadrature of the AMS of each category, is maximized. The AMS of each category is defined as:\\
\begin{linenomath*}
\begin{equation}
\label{eq:AMS_eq}
    \mathrm{AMS}=\sqrt{2\left[(S+B) \ln \left(1+\frac{S}{B}\right)-S\right]}.
\end{equation}
\end{linenomath*}

\begin{figure*}[thb]
  \centering
  \includegraphics[width=0.45\textwidth]{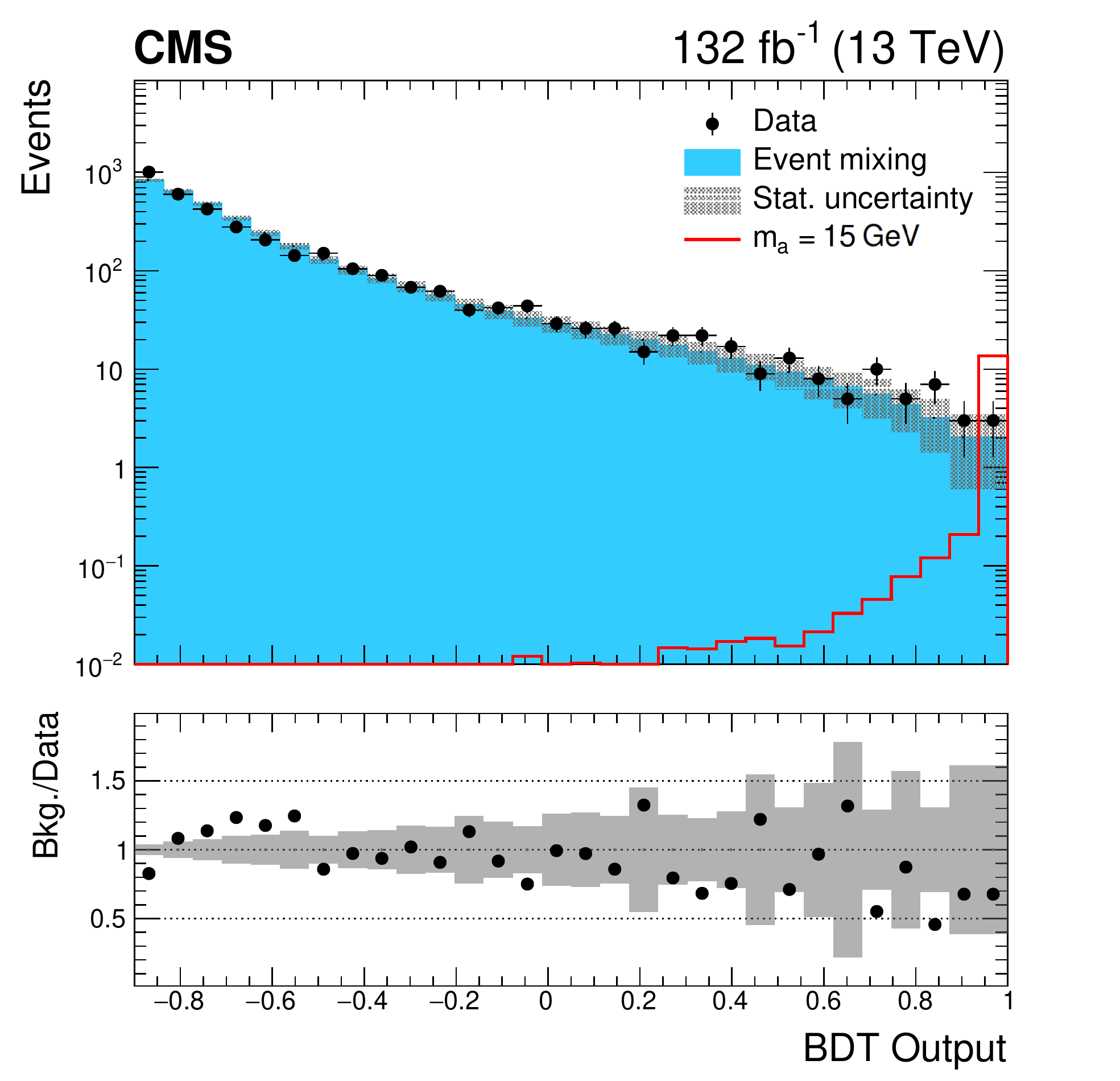}\hfil
  \includegraphics[width=0.45\textwidth]{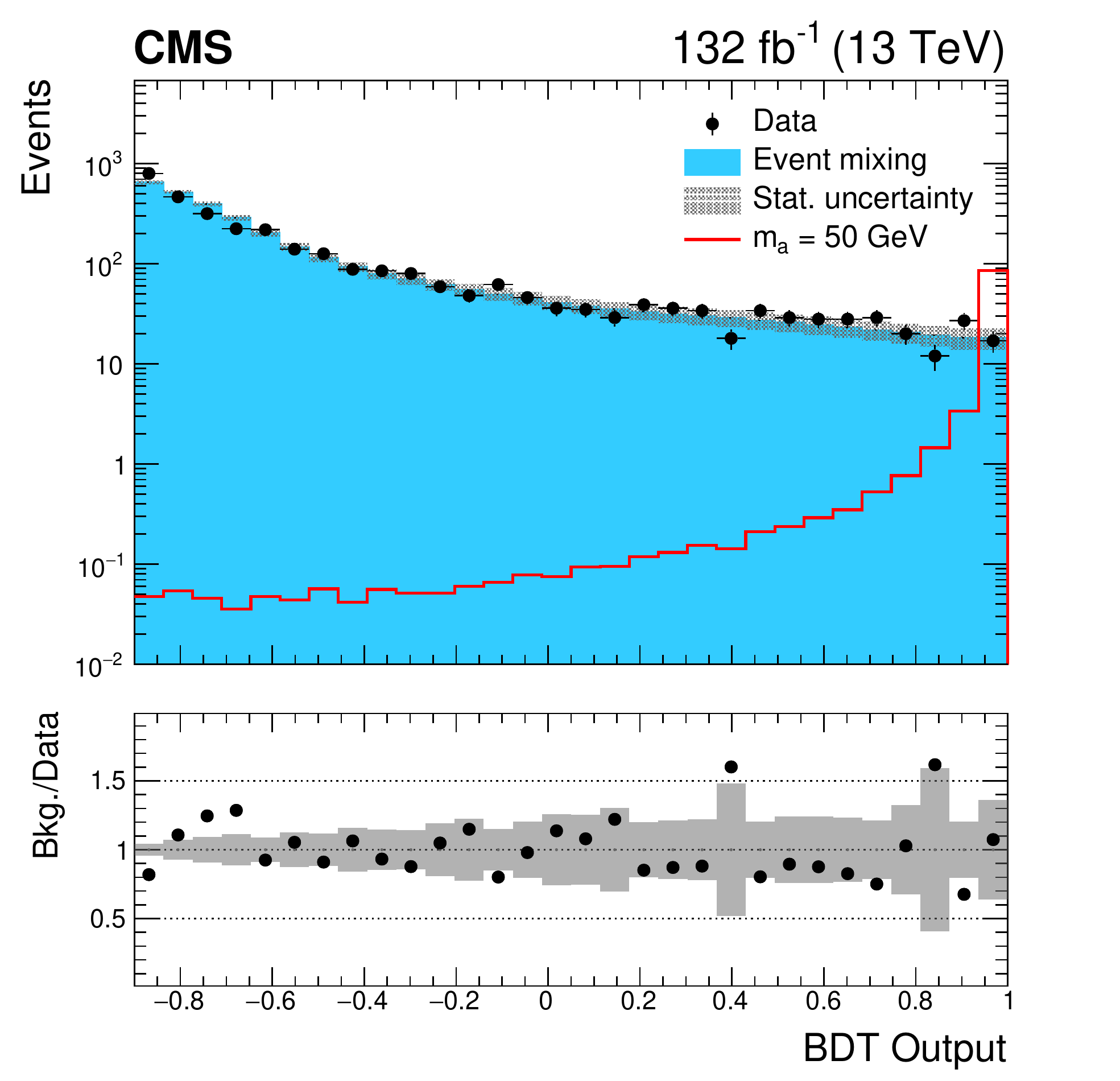}\hfil
  \caption{Distribution of the BDT output for ${\msc = 15}$\GeV (left) and 50\GeV (right) in data and simulated signal and event mixing (after smoothing) events. Events shown are selected after fulfilling the selection criteria described in Section~\ref{sec:EventReconstruction} in the mass window ${110 < \mgggg < 180}$\GeV, while the signal is scaled with a cross-section of 1\unit{pb}.}
  \label{fig:BDTOutput}
\end{figure*}

In Eq.~\ref{eq:AMS_eq}, S refers to the number of expected signal events and B refers to the number of estimated background events. In order to minimize the impact of statistical fluctuations on the optimization of the BDT selection, the output BDT distribution of the event mixing dataset is smoothed, using the super-smoothing technique~\cite{SmmothSuper,SmmothSuper2}, prior to being used in this procedure. The distributions of the BDT output for data, simulated signal events, and event mixing dataset, after smoothing the BDT output distribution, is shown in Fig.~\ref{fig:BDTOutput} for ${\msc = 15}$ and 50\GeV. Events shown in this distribution are selected after passing the criteria described in Section~\ref{sec:EventReconstruction} and are in the mass window ${110 < \mgggg < 180}$\GeV.

A priori, the greater the number of categories, the better the significance is within the uncertainties. Therefore this optimization procedure was performed separately for each of the nominal signal hypotheses for up to five categories of the 4-photon BDT score. However, as an increase of less than 1\% in the AMS value was observed when increasing the number of categories beyond one, only a single category based on the BDT output was created for each pseudoscalar boson mass hypothesis. Table~\ref{tab:sig_table} summarizes the minimum BDT selection and the efficiency of that selection obtained for each of the nominal signal hypotheses. For the intermediate \msc values, the BDT selection obtained from the closest nominal signal hypothesis is applied. For each signal hypothesis, events that pass the selection on the BDT output are used to obtain final results.

\begin{table*} [hbtp]
    \centering
    \topcaption{Summary of the optimized BDT output threshold values and the efficiency with respect to a selection on this output for each of the nominal signal hypothesis.}
        \begin{tabular}{ccc}
         \multirow{2}{*}{\msc (\GeVns{})}& Minimum 4-photon event classification BDT & Signal efficiency \\
                                             & output value  & of the selection on BDT \\
      \hline
      15 & 0.883 & 88\%\\
      20 & 0.891 & 87\%\\
      25 & 0.876 & 86\%\\
      30 & 0.897 & 84\%\\
      35 & 0.931 & 82\%\\
      40 & 0.945 & 78\%\\
      45 & 0.952 & 80\%\\
      50 & 0.958 & 80\%\\
      55 & 0.976 & 77\%\\
      60 & 0.987 & 71\%\\
    \end{tabular}
    \label{tab:sig_table}
\end{table*}

\section{Statistical procedure}
\label{sec:StatProd}

The statistical procedure used in this analysis is identical to that described in Ref.~\cite{likelihoodRatio}, as developed by the ATLAS and CMS Collaborations. Simultaneous unbinned maximum likelihood fits are performed to the \mgggg distributions of all analysis categories, with an \msc granularity of 0.5\GeV for ${15 < \msc < 40}$\GeV and 1\GeV for $40 \leq \msc \leq 62\GeV$. A likelihood function is defined for each analysis category using analytic models to describe the \mgggg distributions of signal and background events with nuisance parameters to account for the experimental and theoretical systematic uncertainties described in Section~\ref{sec:Systematics}. The best fit values and confidence intervals for the parameters of interest are estimated using a profile likelihood test statistic:

\begin{linenomath*}
\begin{equation}
\label{eq:likelihood}
    q(\vec{\alpha}) = - 2\ln \left( \frac{\mathcal{L}(\vec{\alpha},\hat{\vec{\theta}}_{\vec{\alpha}})}{\mathcal{L}(\hat{\vec{\alpha}},\hat{\vec{\theta}})} \right).
\end{equation}
\end{linenomath*}

Where the quantities $\hat{\vec{\alpha}}$ and $\hat{\vec{\theta}}$ describe the unconditional maximum likelihood estimates for the parameters of interest and the nuisance parameters, respectively, whereas $\hat{\vec{\theta}}_{\vec{\alpha}}$ corresponds to the conditional maximum likelihood estimate for fixed values of the parameters of interest, $\vec{\alpha}$. The value of twice the negative logarithm of the likelihood ratio, Eq.~\ref{eq:likelihood} is minimized when a fit of these functions is performed on the \mgggg distribution. A penalty term, equal to the number of parameters in the functions, is added to the \dNLL to prevent the addition of unnecessary floating parameters in the fit.

\section{Signal model}
\label{sec:Signal}

\begin{figure}[!hbtp]
        \centering
        \includegraphics[width=.41\linewidth]{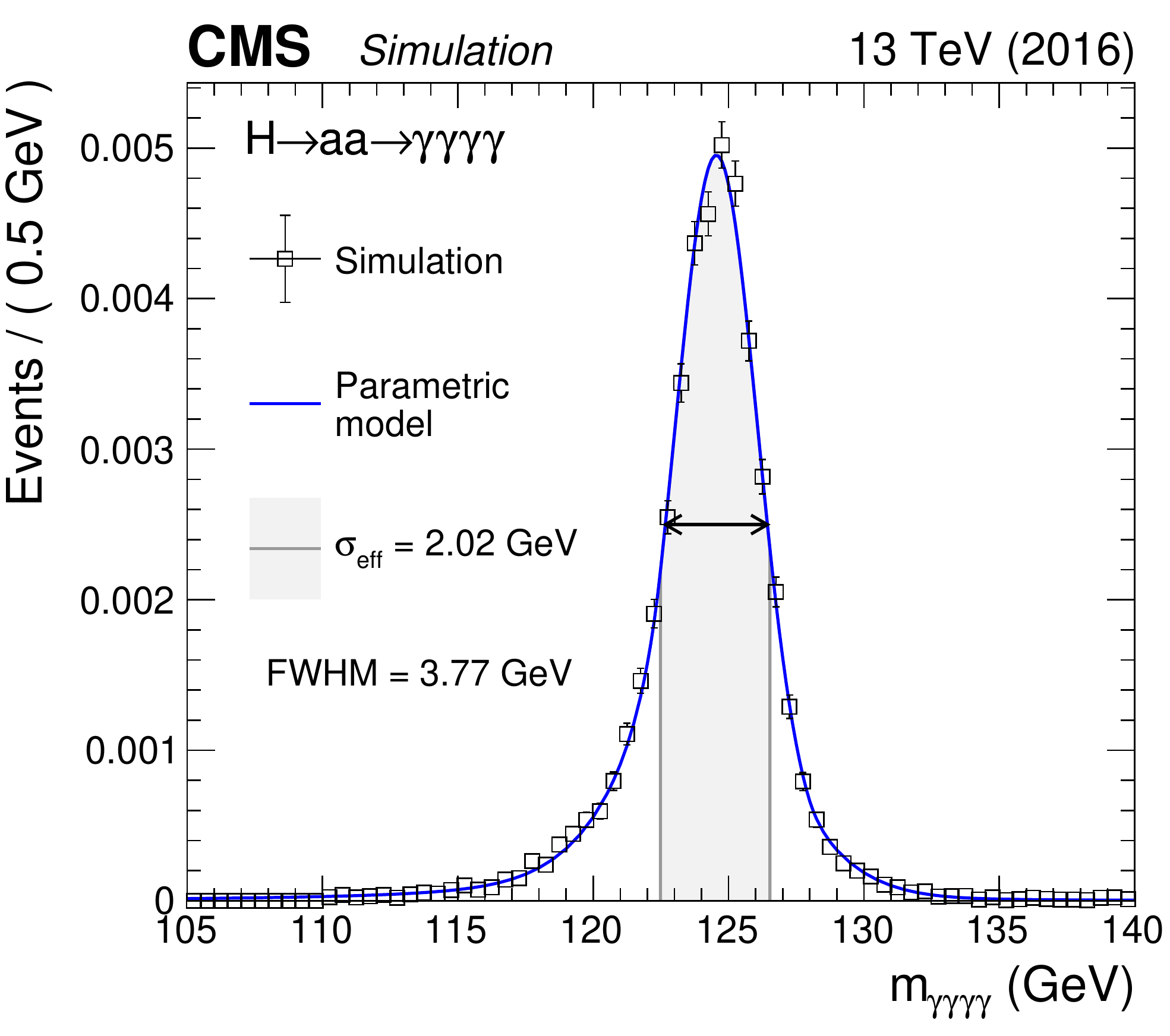}
        \qquad
        \includegraphics[width=.41\linewidth]{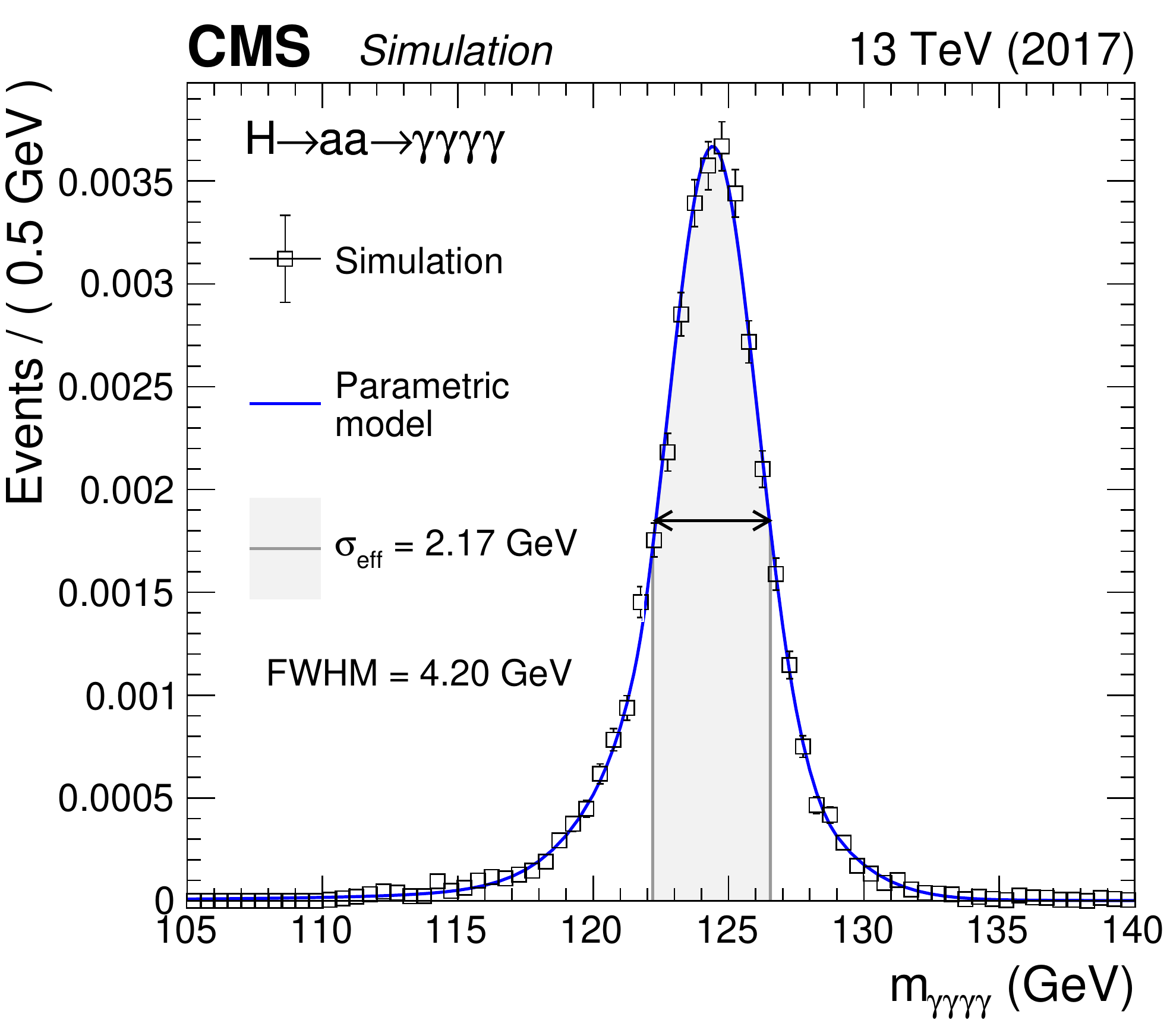}
        \includegraphics[width=.41\linewidth]{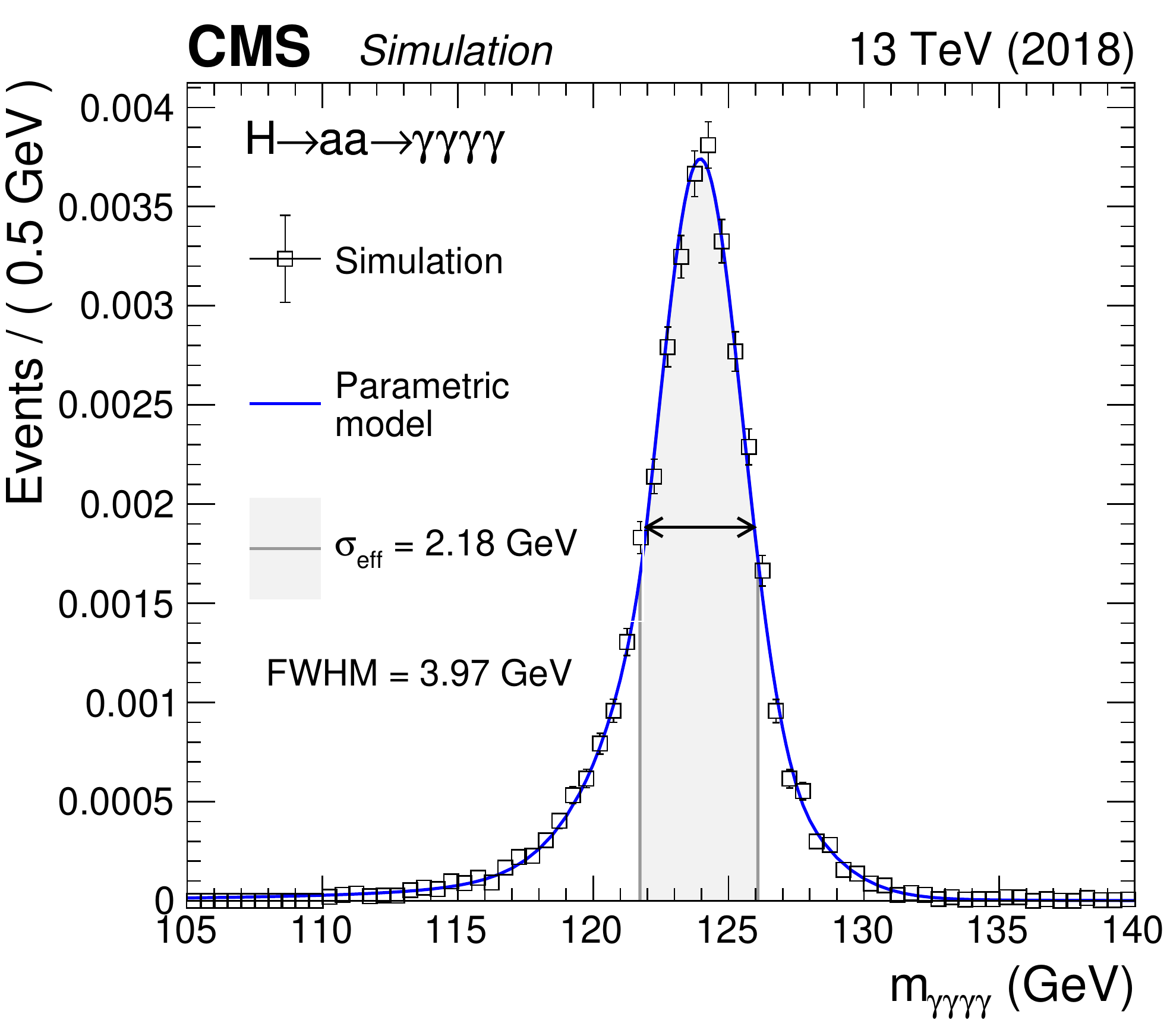}
        \caption{The parameterized signal shape for ${\msc = 15}$\GeV is shown for the 2016 (upper left), 2017 (upper right), and 2018 (lower) data-taking years. Separate signal models are built for each of the three data-taking years, which are then scaled by the appropriate luminosity and summed in order to construct the final signal model. The open squares represent simulated events and the blue line is the corresponding model. Also shown is the $\sigma_{\text{eff}}$ value (half the width of the narrowest interval containing 68.3\% of the invariant mass distribution), with the corresponding interval as a gray band and the FWHM, with the corresponding interval marked with a double arrow.}
        \label{fig:Sig_M15}
\end{figure}

The signal shape for the \mgggg distribution, for each nominal signal hypothesis, is constructed from simulation. After all of the analysis selection criteria are applied, a unique signal model is built for each nominal signal hypothesis for each of the three data-taking years ({2016--2018}). The \mgggg distribution is modeled with a double-sided Crystal Ball (CB) function~\cite{DoubleCrystalBallRef}, which is a modified version of the standard CB function with two independent power-low tails.
These signal models, scaled by the integrated luminosity for each year, are summed in order to construct the final model. The signal models for each year are shown in Fig.~\ref{fig:Sig_M15} for ${\msc = 15}$\GeV. The full width at half maximum (FWHM) and the effective standard deviation ($\sigma_{\text{eff}}$), defined as half the width of the smallest interval containing 68\% of the \mgggg distribution, are also shown.

Two factors need to be considered to build the signal models for the intermediate mass hypotheses: the shape of the \mgggg distribution and its yield. Since the shape of the \mgggg distribution is not found to change significantly within a 5\GeV window around the nominal mass hypothesis, only the yield of the signal model is parameterized as a function of \msc for the intermediate mass hypotheses. For each intermediate point, a signal model is constructed from the \mgggg shape of the nearest nominal mass hypothesis and the interpolated yield between the two nearest nominal mass hypotheses.

\section{Background model}
\label{sec:Background}
The background model is built to describe the shape of the \mgggg distribution that results from processes other than the signal process. Since the shape of this distribution is not known, different functional forms must be considered in the construction of the background model. The function chosen to describe the background can result in a different number of estimated events in the signal peak and, as a result, affect the measured signal strength. This inherent uncertainty associated with choosing a background function is incorporated into the statistical uncertainty from the fit model via the discrete profiling method~\cite{DiscreteProfilingMethod}. This method describes background modeling performed using data as implemented in Ref.~\cite{Khachatryan:2014ira}, and treats the choice of the background function as a discrete nuisance parameter in the likelihood fit to data.

As part of the background modeling procedure, the candidate functions considered in the fit are exponentials, Bernstein polynomials, Laurent series, and power law functions. A subset of functions from each family are used to build the background model. For each family, the maximum order of parameters is fixed by means of an F-test~\cite{ftest}, and the minimum order is determined by applying a requirement on the goodness-of-fit. A penalty is added to \dNLL to take into account the number of floating parameters in each candidate function. When making a measurement of a given parameter of interest, the discrete profiling method minimizes the overall \dNLL considering all allowed functions. 

The fits are performed over the range $110 < \mgggg < 180$\GeV, and the data from the three years (2016--2018) are combined in order to construct the background model. A unique background model is created corresponding to each \msc hypothesis. For each \msc hypothesis, an ensemble of pseudo-experiments was generated using the various background functions. Each pseudo-experiment was fitted using the discrete profiling method, and it was established that the chosen functional form, used to describe the background, does not introduce any potential bias in the signal strength measurement. 

\section{Systematic uncertainties}
\label{sec:Systematics}

The systematic uncertainty associated with the background estimation is taken into account by the discrete profiling method, as described in Section\ref{sec:Background}. There are two kinds of systematic uncertainties that affect the signal model: those that modify the shape of the \mgggg distribution, and those that leave the shape of the \mgggg distribution unchanged but affect the overall normalization of the signal model.

The uncertainties that affect the shape of the \mgggg distribution, which are incorporated in the signal model as nuisance parameters, are described below. These uncertainties are typically related to the energy of the individual photons, and they affect the mean and width of the signal model.
\begin{enumerate}
    \item \emph{Photon energy scale and resolution}: Corrections are applied to the photon energy scale in data and to the energy resolution in simulation. The uncertainties related to these corrections are computed using \Zee events~\cite{CMS:2020ume}. The resulting uncertainty in the energy scale is {0.05--0.15}\%, depending on the photon \pt.
    \item \emph{Nonlinearity of the photon energy scale}: Any remaining differences in the linearity of the photon energy scale between data and simulation are covered by this uncertainty, which is estimated using Lorentz-boosted \Zee events. The procedure for estimating this uncertainty is detailed in Ref.~\cite{Hgg_properties_13_v2}. An uncertainty of 0.1\% on the photon energy scale is assigned in this analysis, which accounts for the nonlinearity across the full range of photon \pt values.
    \item \emph{Shower shape corrections}: This uncertainty is associated with the imperfect modeling of shower shapes in simulation, and it is estimated by comparing the energy scale before and after any corrections are applied to the shower shape variables as described in Ref.~\cite{CMS:2020ume}. This uncertainty in the energy scale is at most {0.01--0.15}\%, and it is dependent on the photon shower-shape and position in the detector. 
    \item \emph{Nonuniformity of light collection}: Within a given ECAL crystal, there is an uncertainty associated with the modeling of the light collection as a function of the emission depth. This uncertainty is estimated by comparing simulation with the longitudinal shower profile estimates, and the procedure is detailed in Ref.~\cite{Hgg_properties_13_v2}. The magnitude of this uncertainty is 0.{07--0.25}\%, depending on the photon shower-shape.
    \item \emph{Modeling of material in front of the ECAL}: The behavior of electromagnetic showers is affected by the amount of material present in front of the ECAL. This behavior may not be well modeled in simulation, and thus special samples with variations in the amount of upstream material are used to compute the impact on the photon energy scale~\cite{Khachatryan:2014ira}. For most central photons, the magnitude of this uncertainty ranges {0.02--0.05}\% and increases to approximately 0.24\% for photons in the endcaps.
\end{enumerate}

The uncertainties that affect the normalization of the signal model are listed below.
  \begin{enumerate}
    \item \emph{Integrated luminosity}: Uncertainties in the luminosity measurement are estimated to be 1.2\% (2016), 2.3\% (2017), and 2.5\% (2018) by CMS luminosity monitoring~\cite{CMS-PAS-LUM-17-001,CMS-PAS-LUM-17-004,CMS-PAS-LUM-18-002}. The uncertainty in the total integrated luminosity of the three years together is 1.6\%. The uncertainties for each dataset are partially correlated in order to account for the common sources in the luminosity measurement schemes.
    \item \emph{Photon identification BDT score:} The systematic uncertainty caused by the imperfect simulation of the input variables that are used to train the photon identification MVA is estimated by the procedure described in Ref.~\cite{CMS:2020ume}. The average magnitude of the resulting uncertainty is below 0.25\% across the full \msc range.
    \item \emph{Trigger efficiency}: The efficiency of the trigger selection is measured using a ``tag-and-probe" technique on \Zee events~\cite{CMS:2011aa}. An additional uncertainty is introduced to account for a gradual shift in the timing to the inputs of the ECAL's hardware level trigger in the region ${\abs{\eta} > 2.0}$, which caused a specific trigger inefficiency during 2016-2017 data-taking~\cite{Sirunyan:2020zal}. The size of this uncertainty across the \msc range is around 0.5\% for 2016 and 2018 data-taking and around 1.5\% for 2017 data-taking.
    \item \emph{Photon preselections}: The systematic uncertainty on photon-based preselection is computed as the uncertainty on the ratio between the efficiency measured in data and in simulation. This is measured with the tag-and-probe technique using \Zee events. The average magnitude of this uncertainty across the \msc range is about 5\%.
  \end{enumerate}

A summary of all the systematic uncertainties considered in this analysis is given in Table~\ref{tab:systematics}. The impact of systematic uncertainties on the expected limit is about 1\% across the \msc range, and so the analysis sensitivity is primarily limited by the expected signal event yields. Because of this, only the main sources of systematic uncertainties are taken into account in the final result.

\begin{table*} [hbtp]
    \centering
    \topcaption{Summary of the systematic uncertainties considered in this analysis.}
        \begin{tabular}{lc}
        $\mgggg$ distribution shape &  {2016--2018} \\ \hline
        Photon energy scale and resolution  &{0.05--0.15}\%  \\
        Nonlinearity of the photon energy scale &  0.10\%  \\
        Shower shape corrections &  {0.01--0.15}\%  \\ 
        Nonuniformity of light collection  &{0.07--0.25}\% \\ 
        Modeling of material in front of the ECAL &  {0.02--0.05}\% (EB) and 0.24\% (EE) \\
        \end{tabular}
        \vskip\cmsTabSkip
        \begin{tabular}{lccc}
        Signal model normalization & 2016 & 2017 & 2018 \\\hline
        Integrated luminosity & 1.20\% & 2.30\% & 2.50\% \\
        Photon identification & 0.25\% & 0.25\% & 0.25\% \\
        Trigger efficiency & 0.50\% & 1.50\% & 0.50\% \\ 
        Photon preselections & 5.00\% & 5.00\% & 5.00\% \\
        \end{tabular}
    \label{tab:systematics}
\end{table*}

\section{Results}
\label{sec:Results}

\begin{figure}[htbp]
 \centering
\includegraphics[width=0.45\textwidth]{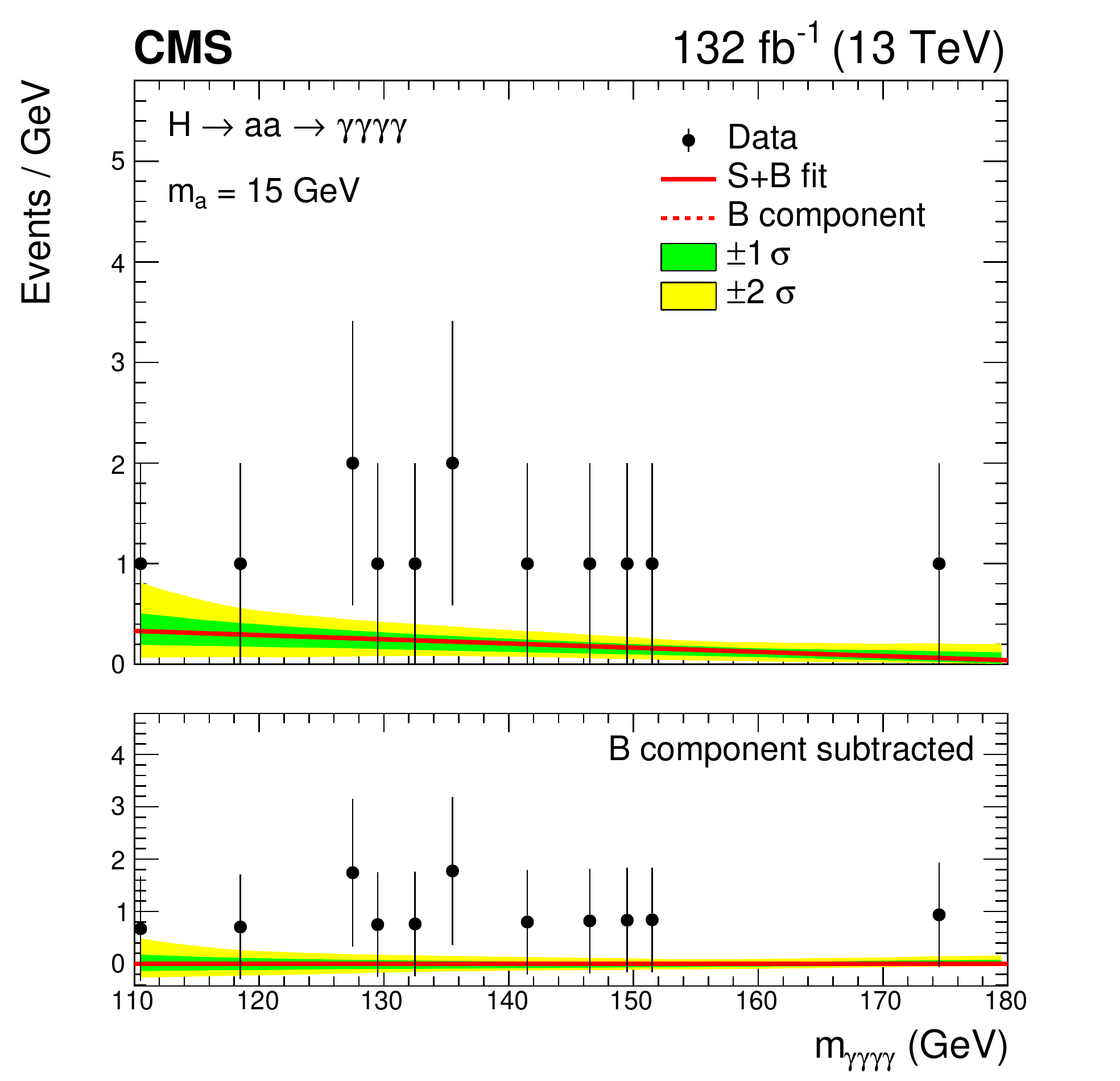}
\includegraphics[width=0.45\textwidth]{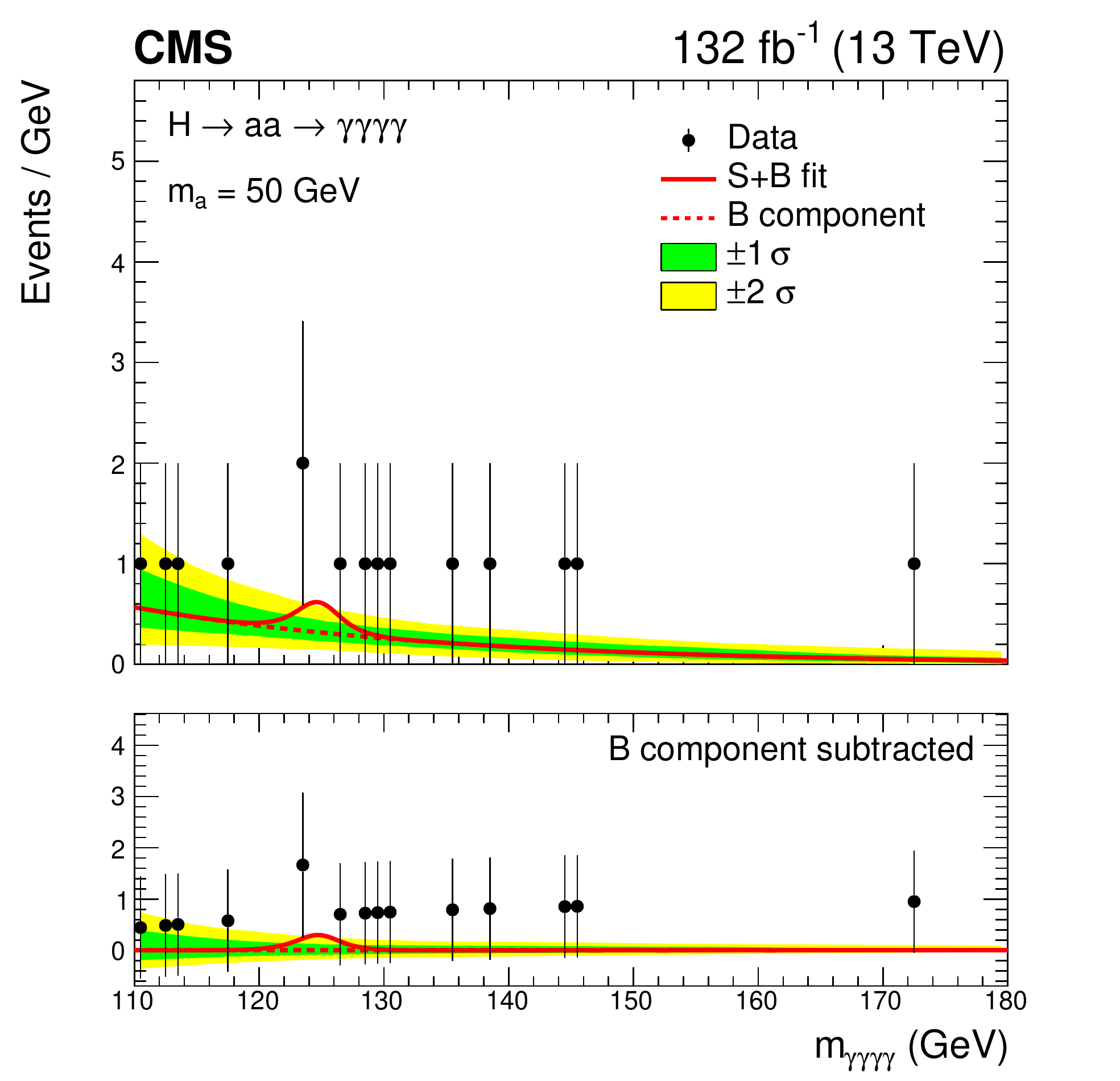}
  \caption{The invariant mass distribution, \mgggg, for
data (black points) and the signal-plus-background model fit is shown for ${\msc = 15}$\GeV (left) and ${\msc = 50}$\GeV (right). The solid red line shows the
total signal-plus-background contribution, whereas the dashed red line shows the background
component only. The lower panel in each plot shows the residual signal yield after the background subtraction. The one (green) and two (yellow) standard deviation bands include the uncertainties in the background component of the fit.}
 \label{fig:BkgModel}
\end{figure}

The data collected by the CMS experiment in 2016, 2017, and 2018 are combined for the fit. The data and the signal-plus-background model that was fit to the \mgggg distribution are shown in Fig.~\ref{fig:BkgModel} for $\msc = 15$ and 50\GeV.

\begin{figure}[!hbt]%
        \centering
        \includegraphics[width=.60\linewidth]{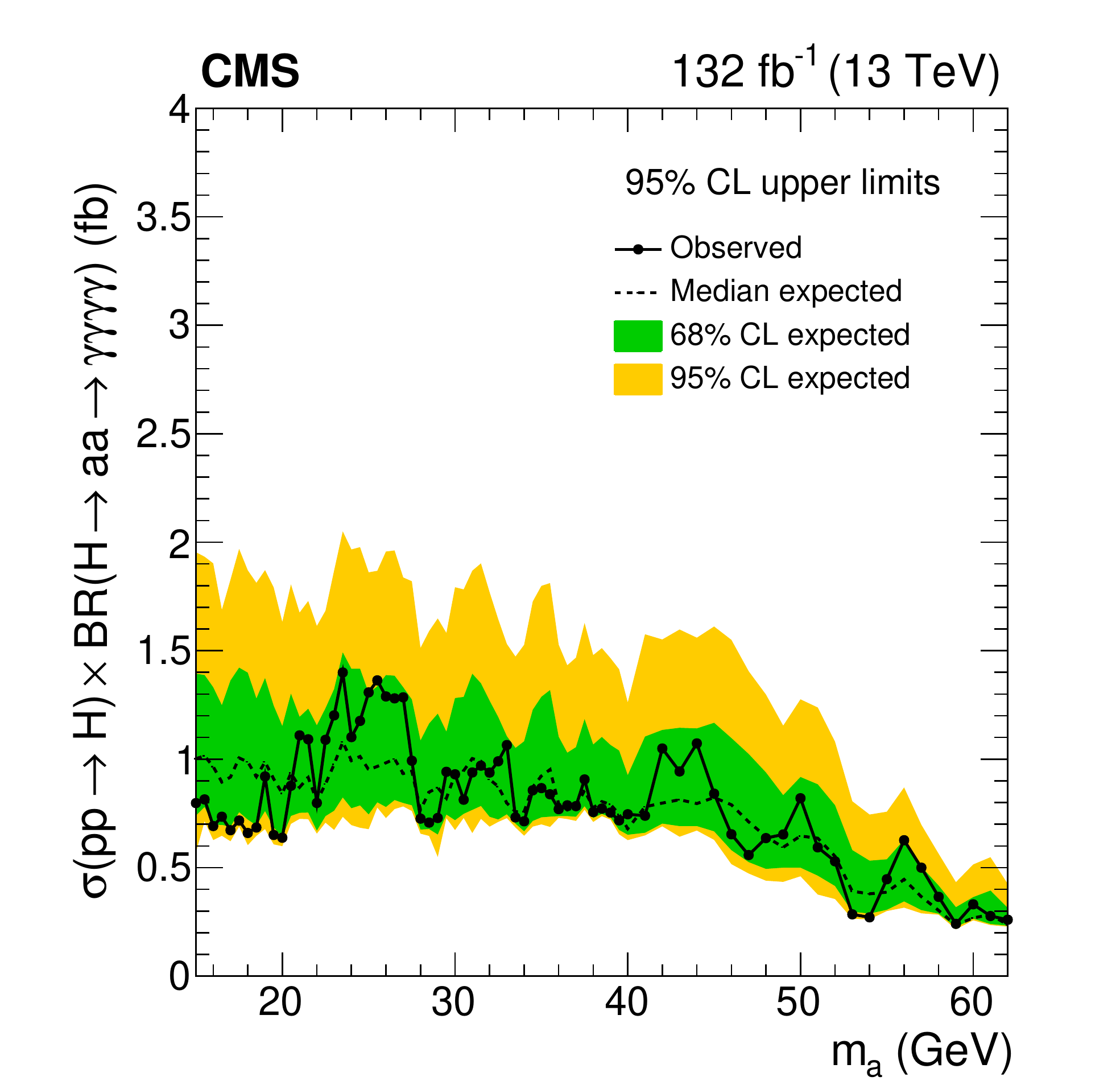}
        \caption{Expected and observed 95\% \CL limits on the product of the production cross section of the Higgs boson and the branching fraction into four photons via a pair of pseuodscalars, $\sigmaH\,\BR$, are shown as a function of \msc. The green (yellow) bands represent the 68\% (95\%) expected limit \CL intervals. The fluctuation between individual points is due to the statistical limitation of the data sample and the result of individual BDT training networks utilized for each individual mass point scenario.}%
        \label{fig:Limit}
\end{figure}

{\tolerance=800
No significant deviation from the background-only hypothesis is observed. Upper limits are set at the 95\% confidence level (\CL) on the product of the production cross section of the Higgs boson and the branching fraction into four photons via a pair of pseuodscalars, $\sigmaH\,\BR$. This is done using the modified frequentist approach for \CLs, with the LHC profile likelihood ratio used as the test statistic~\cite{HybridNew,likelihoodRatio}. The observed (expected) limit, shown in Fig.~\ref{fig:Limit}, ranges from 0.80 (1.00)\unit{fb}for $\msc = 15$\GeV to 0.26 (0.24)\unit{fb} for ${\msc = 62}$\GeV. For comparison, the Higgs production cross section for all channels combined is 52\unit{pb}~\cite{deFlorian:2016spz}.
\par}

The results presented in this section are provided in a tabulated form in the HEPDATA record~\cite{hepdata} for this analysis.


\section{Summary}
\label{sec:Summary}

A search for a pair of light pseudoscalar bosons produced from the decay of the 125\GeV Higgs boson, which subsequently decay into photons, is presented. The analysis is based on proton-proton collision data collected at \sqrts by the CMS experiment at the LHC in 2016, 2017, and 2018, which corresponds to a total integrated luminosity of 132\fbinv. The analysis probes pseudoscalar bosons ranging in mass (\msc) from 15 to 62\GeV. No significant deviation from the background-only hypothesis is observed. Upper limits are set at 95\% confidence level on the product of the production cross section of the Higgs boson and the branching fraction into four photons via a pair of pseuodscalars, ${\sigmaH\,\BR}$. The observed (expected) limits range from 0.80 (1.00)\unit{fb} for ${\msc = 15}$\GeV to 0.26 (0.24)\unit{fb} for ${\msc = 62}$\GeV.

\begin{acknowledgments}
  We congratulate our colleagues in the CERN accelerator departments for the excellent performance of the LHC and thank the technical and administrative staffs at CERN and at other CMS institutes for their contributions to the success of the CMS effort. In addition, we gratefully acknowledge the computing centers and personnel of the Worldwide LHC Computing Grid and other centers for delivering so effectively the computing infrastructure essential to our analyses. Finally, we acknowledge the enduring support for the construction and operation of the LHC, the CMS detector, and the supporting computing infrastructure provided by the following funding agencies: BMBWF and FWF (Austria); FNRS and FWO (Belgium); CNPq, CAPES, FAPERJ, FAPERGS, and FAPESP (Brazil); MES and BNSF (Bulgaria); CERN; CAS, MoST, and NSFC (China); MINCIENCIAS (Colombia); MSES and CSF (Croatia); RIF (Cyprus); SENESCYT (Ecuador); MoER, ERC PUT and ERDF (Estonia); Academy of Finland, MEC, and HIP (Finland); CEA and CNRS/IN2P3 (France); BMBF, DFG, and HGF (Germany); GSRI (Greece); NKFIH (Hungary); DAE and DST (India); IPM (Iran); SFI (Ireland); INFN (Italy); MSIP and NRF (Republic of Korea); MES (Latvia); LAS (Lithuania); MOE and UM (Malaysia); BUAP, CINVESTAV, CONACYT, LNS, SEP, and UASLP-FAI (Mexico); MOS (Montenegro); MBIE (New Zealand); PAEC (Pakistan); MES and NSC (Poland); FCT (Portugal); MESTD (Serbia); MCIN/AEI and PCTI (Spain); MOSTR (Sri Lanka); Swiss Funding Agencies (Switzerland); MST (Taipei); MHESI and NSTDA (Thailand); TUBITAK and TENMAK (Turkey); NASU (Ukraine); STFC (United Kingdom); DOE and NSF (USA).
   
  \hyphenation{Rachada-pisek} Individuals have received support from the Marie-Curie program and the European Research Council and Horizon 2020 Grant, contract Nos.\ 675440, 724704, 752730, 758316, 765710, 824093, 884104, and COST Action CA16108 (European Union); the Leventis Foundation; the Alfred P.\ Sloan Foundation; the Alexander von Humboldt Foundation; the Belgian Federal Science Policy Office; the Fonds pour la Formation \`a la Recherche dans l'Industrie et dans l'Agriculture (FRIA-Belgium); the Agentschap voor Innovatie door Wetenschap en Technologie (IWT-Belgium); the F.R.S.-FNRS and FWO (Belgium) under the ``Excellence of Science -- EOS" -- be.h project n.\ 30820817; the Beijing Municipal Science \& Technology Commission, No. Z191100007219010; the Ministry of Education, Youth and Sports (MEYS) of the Czech Republic; the Hellenic Foundation for Research and Innovation (HFRI), Project Number 2288 (Greece); the Deutsche Forschungsgemeinschaft (DFG), under Germany's Excellence Strategy -- EXC 2121 ``Quantum Universe" -- 390833306, and under project number 400140256 - GRK2497; the Hungarian Academy of Sciences, the New National Excellence Program - \'UNKP, the NKFIH research grants K 124845, K 124850, K 128713, K 128786, K 129058, K 131991, K 133046, K 138136, K 143460, K 143477, 2020-2.2.1-ED-2021-00181, and TKP2021-NKTA-64 (Hungary); the Council of Science and Industrial Research, India; the Latvian Council of Science; the Ministry of Education and Science, project no. 2022/WK/14, and the National Science Center, contracts Opus 2021/41/B/ST2/01369 and 2021/43/B/ST2/01552 (Poland); the Funda\c{c}\~ao para a Ci\^encia e a Tecnologia, grant CEECIND/01334/2018 (Portugal); the National Priorities Research Program by Qatar National Research Fund;  MCIN/AEI/10.13039/501100011033, ERDF ``a way of making Europe", and the Programa Estatal de Fomento de la Investigaci{\'o}n Cient{\'i}fica y T{\'e}cnica de Excelencia Mar\'{\i}a de Maeztu, grant MDM-2017-0765 and Programa Severo Ochoa del Principado de Asturias (Spain); the Chulalongkorn Academic into Its 2nd Century Project Advancement Project, and the National Science, Research and Innovation Fund via the Program Management Unit for Human Resources \& Institutional Development, Research and Innovation, grant B05F650021 (Thailand); the Kavli Foundation; the Nvidia Corporation; the SuperMicro Corporation; the Welch Foundation, contract C-1845; and the Weston Havens Foundation (USA).
\end{acknowledgments}

\bibliography{auto_generated} 

\cleardoublepage \appendix\section{The CMS Collaboration \label{app:collab}}\begin{sloppypar}\hyphenpenalty=5000\widowpenalty=500\clubpenalty=5000
\cmsinstitute{Yerevan Physics Institute, Yerevan, Armenia}
{\tolerance=6000
A.~Tumasyan\cmsAuthorMark{1}\cmsorcid{0009-0000-0684-6742}
\par}
\cmsinstitute{Institut f\"{u}r Hochenergiephysik, Vienna, Austria}
{\tolerance=6000
W.~Adam\cmsorcid{0000-0001-9099-4341}, J.W.~Andrejkovic, T.~Bergauer\cmsorcid{0000-0002-5786-0293}, S.~Chatterjee\cmsorcid{0000-0003-2660-0349}, K.~Damanakis\cmsorcid{0000-0001-5389-2872}, M.~Dragicevic\cmsorcid{0000-0003-1967-6783}, A.~Escalante~Del~Valle\cmsorcid{0000-0002-9702-6359}, R.~Fr\"{u}hwirth\cmsAuthorMark{2}\cmsorcid{0000-0002-0054-3369}, M.~Jeitler\cmsAuthorMark{2}\cmsorcid{0000-0002-5141-9560}, N.~Krammer\cmsorcid{0000-0002-0548-0985}, L.~Lechner\cmsorcid{0000-0002-3065-1141}, D.~Liko\cmsorcid{0000-0002-3380-473X}, I.~Mikulec\cmsorcid{0000-0003-0385-2746}, P.~Paulitsch, F.M.~Pitters, J.~Schieck\cmsAuthorMark{2}\cmsorcid{0000-0002-1058-8093}, R.~Sch\"{o}fbeck\cmsorcid{0000-0002-2332-8784}, D.~Schwarz\cmsorcid{0000-0002-3821-7331}, S.~Templ\cmsorcid{0000-0003-3137-5692}, W.~Waltenberger\cmsorcid{0000-0002-6215-7228}, C.-E.~Wulz\cmsAuthorMark{2}\cmsorcid{0000-0001-9226-5812}
% ...
\par}
\vskip\cmsinstskip
\dag:~Deceased\\
$^{1}$Also at Yerevan State University, Yerevan, Armenia\\
$^{2}$Also at TU Wien, Vienna, Austria\\
% ...
\end{sloppypar}
\end{document}